\documentclass{article}
\usepackage{fortschritte}
\def\calN={{\cal N}}
\def\nn{\nonumber}

\def\ttt{{\tilde t}}

\def\tq{{\tilde q}}
\def\hm{{\widehat m}}

\def\ie{{\it i.e.}}
\def\half{{{1 \over 2}}}

\def\pmb#1{\setbox0=\hbox{#1}%
 \kern-.025em\copy0\kern-\wd0
 \kern.05em\copy0\kern-\wd0
 \kern-.025em\raise.0433em\box0 }
\font\cmss=cmss10
\font\cmsss=cmss10 at 7pt
\def\rlx{\relax\leavevmode}
\def\Cop{\relax\,\hbox{$\kern-.3em{\rm C}$}}
\def\Rop{\relax{\rm I\kern-.18em R}}
\def\Nop{\relax{\rm I\kern-.18em N}}
\def\Pop{\relax{\rm I\kern-.18em P}}

\def\Zop{\rlx\leavevmode\ifmmode\mathchoice{\hbox{\cmss Z\kern-.4em Z}}
 {\hbox{\cmss Z\kern-.4em Z}}{\lower.9pt\hbox{\cmsss Z\kern-.36em Z}}
 {\lower1.2pt\hbox{\cmsss Z\kern-.36em Z}}\else{\cmss Z\kern-.4em
 Z}\fi}

\def\m{{m}}

\def\ads{{AdS$_5\times$ S$^5$}}

\def\bbbone {{\mathchoice {\rm 1\mskip-4mu l} {\rm 1\mskip-4mu l}
{\rm 1\mskip-4.5mu l} {\rm 1\mskip-5mu l}}}
                                         
\def\beq{\begin{equation}}                     %
\def\eeq{\end{equation}}                       %
\def\bea{\begin{eqnarray}}                     
\def\eea{\end{eqnarray}}                       
\begin {document}                 
\def\titleline{
%
%
%
Strings and branes in plane waves             
}
\def\email_speaker{
{\tt 
%
%
 mrg@phys.ethz.ch                            
}}
\def\authors{
%
%
%
%
%
Matthias R. Gaberdiel                        
}
\def\addresses{
%
%
%
%
Institute for Theoretical Physics \\
ETH H\"onggerberg\\
CH-8093 Z\"urich, Switzerland\sp
                                            %
                                            %
                                            %
}
\def\abstracttext{
%
%
An overview of string theory in the maximally          
supersymmetric plane-wave                
background is given, and some supersymmetric
D-branes are discussed.
}
\large
\makefront

\section{Introduction \& Motivation}

According to the Maldacena conjecture \cite{maldacena} type IIB string
theory on \ads\ is dual to four-dimensional ${\cal N}=4$
super Yang-Mills theory with gauge group SU$(N)$,  
where the parameters of the two theories are related as 
\bea
g_s & = & g_{}^2 \,,\nn\\
R & = & (g_{}^2 N)^{1/4} \,.
\eea
Here $g_s$ is the string coupling constant, while $g_{}$ denotes the
coupling constant of the Yang-Mills theory. The second equation
implies that the radius of \ads\ is proportional to a power of the
t'Hooft parameter $x=g_{}^2 N$. In the large $N$ limit for fixed but
large t'Hooft parameter $x$, the string theory is weakly coupled and   
the curvature terms are small. We can then use the supergravity
approximation to string theory to calculate quantities on the
string theory side, and compare them with Yang-Mills theory.
This is the limit in which most tests of the
Maldacena conjecture have been performed. However, the above duality
is meant to hold not only in this limit, but also as a 
duality at {\it finite} $N$, relating a string theory with a
four-dimensional conformal field theory. Unfortunately,  
string theory on \ads\ is so far not solvable, and it is therefore
difficult to check this conjecture directly.

Recently it was observed that there exists a limit in which the
string theory becomes exactly solvable. This is the so-called 
{\it Penrose limit} \cite{maldetal,met,mt,hulletal0,hulletal}. One can 
take a corresponding limit in the dual gauge theory, and this then
allows for a quantitative analysis of at least  some aspects of the
Maldacena correspondence. This line of thought has been successfully
pursued in a number of papers (see for example \cite{sv,kris,const,sz} 
for a very incomplete list of  papers, and 
\cite{ari,plefka,sv1,ssj} for some recent reviews). These developments
will be described by other speakers; here we shall concentrate on
explaining how the string theory can be exactly solved in this
background.  

To begin with we need to explain how to construct the Penrose limit in
detail. To this end recall that the metric of \ads\  can be written as
\beq
ds^2 = R^2 \left[ - \cosh^2 \rho \, dt^2 + d\rho^2 + \sinh^2\rho \, 
d\Omega_3^2
+ \cos^2\theta \, d\psi^2 + d\theta^2 + \sin^2\theta \, {d\Omega_3'}^2 
\right]\,,
\eeq
where the first three terms describe the metric of AdS$_5$, while the
last three terms describe that of S$^5$. Let us consider a null
geodesic that is defined by  
\beq
\rho=0\,,\qquad \theta=0\,,\qquad \psi=t \,.
\eeq
The Penrose limit  is simply the
space that is obtained by blowing up a small neighbourhood of this
null geodesic. To this end introduce light-cone  coordinates
$\tilde{x}^\pm = \pm (t\pm \psi)/\sqrt{2}$, and rescale the coordinates by
defining 
\beq
x^+=\tilde{x}^+\,, \qquad 
x^-=R^2 \tilde{x}^- \,, \qquad
\rho={r\over R}\,, \qquad
\theta = {y\over R}\,,
\eeq
and taking $R\rightarrow\infty$. It is an easy exercise to check that
the above metric becomes in this limit 
\beq
ds^2 = 2\, dx^+ dx^- - {1\over 4}\, (r^2 + y^2)\, dx^+ dx^+ 
+ d{\bf y}^2 + d{\bf r}^2 \,,
\eeq
where ${\bf r}$ and ${\bf y}$ are four-dimensional vectors 
whose length is $r$ and $y$, and whose angular directions are described 
by the three-spheres of AdS$_5$ and S$^5$, respectively. [In the following
we shall also sometimes denote the four coordinates 
${\bf r}$ that come from AdS$_5$ by
$x^1,\ldots,x^4$, and the four coordinates 
${\bf  y}$ that come from  S$^5$ by
$x^5,\ldots,x^8$.]  In addition, \ads\ has a non-trivial RR 5-form
field strength, and in the above limit only the  
\beq
F_{+1234}=F_{+5678}=2\mu
\eeq
components survive. With this normalisation the above metric is then
\beq
ds^2 = 2 dx^+ dx^- - \mu^2 x^I x^I dx^+ dx^+ + dx^I dx^I \,,
\eeq
where $I$ runs from $I=1,\ldots,8$. This is the metric of a 
{\it plane-wave background}. [A general plane-wave background is of
the form 
\beq
ds^2 = 2 dx^+ dx^- + A_{IJ}(x^+) x^I x^J dx^+ dx^+ + dx^I dx^I \,;
\eeq
the above is the special case for which 
$A_{IJ}(x^+)=-\mu^2 \delta_{IJ}$.] In the following 
we shall concentrate on describing string theory in this special case;
it has now been realised that more general backgrounds also lead to
exactly solvable string theories (see for example
\cite{tr,prt,blauetal}). There are also some related backgrounds with
NS-NS flux that can be studied in detail (see for example
\cite{kiritsis}).

\section{String theory in the plane wave background}

It has been known for some time that plane wave backgrounds are exact
solutions of string theory \cite{ak,hs}. However, it has only recently
been realised that string theory is exactly solvable in these
backgrounds in the GS formalism in light-cone gauge \cite{met,mt}. By
varying the sigma-model action with respect to $x^-$, one finds that
$x^+$ satisfies the usual wave equation, which can therefore be solved
by setting  
\beq
x^+ = 2\pi \alpha' p^+ \tau \,,
\eeq
where $\tau$ is the world-sheet time. In light-cone gauge, $x^-$ is
determined by the constraint equations in terms of the transverse
coordinates (and $p^+$), and the Lagrangian for the transverse
coordinates becomes (with $\alpha'=1$) 
\beq
{\cal L} = {1\over 4\pi} \left( \partial_+ x^I \partial_- x^I
- m^2 (x^I)^2  \right)
+ {i \over 2\pi}\left(S^a\partial_+ S^a + \tilde S^a \partial_- \tilde S^a
- 2m\, S^a\, \Pi_{ab} \, \tilde S^b \right)\, .
\label{lcact}
\eeq
Here $m$ is defined by $m=2 \pi p^+ \mu$, and  $S^a$ and $\tilde S^a$ are
$SO(8)$ spinors of the same chirality. The matrix $\Pi$ is given as   
$\Pi = \gamma^1 \gamma^2\gamma^3 \gamma^4$, where the  
$8\times 8$ matrices,
$\gamma^I_{a\dot b}$ and $\gamma^I_{\dot ab}$, are the
off-diagonal  blocks of the $16\times 16$ $SO(8)$ $\gamma$-matrices
and couple $SO(8)$ spinors of opposite chirality.  The presence of
$\Pi$ in the fermionic sector of the Lagrangian (that reflects the RR
5-form flux) breaks the symmetry from $SO(8)$ to $SO(4)\times SO(4)$.

This Lagrangian describes a field theory of eight free massive bosons
and fermions. The mass term implies, in particular, that the theory is
not conformally invariant in light cone gauge.\footnote{It is believed 
\cite{Berkovits} that the covariant  theory is conformally
invariant. By going to light-cone gauge we have used up the conformal
symmetry, and there is {\it a priori} no reason why the resulting
theory should still be conformally invariant. The fact that for flat
Minkowski space the theory in light-cone gauge is still conformally
invariant is a result of some additional symmetries of the Minkowski
background.}  However, it is still a free theory, and we can therefore
solve it explicitly. For example, the equations of motion for the
transverse bosons imply that we have the mode expansion
\beq 
x^I(\sigma,\tau) = \cos(m\tau) x_0^I + {\sin(m\tau)\over m} p_0^I 
+ i \sum_{n\ne 0} {1\over \omega_n} \left(
e^{-i\omega_n \tau + 2\pi n i \sigma} \alpha^I_n 
+ e^{-i\omega_n \tau - 2\pi n i \sigma} \tilde\alpha^I_n \right)\,,
\eeq
where
\beq
\omega_n = \hbox{sign} (n) \sqrt{m^2+n^2} \,.
\eeq
Upon quantisation this leads to the commutation relations
\beq
[\alpha^I_k,\alpha^J_l]  = \omega_k \, \delta^{IJ}\, \delta_{k,-l}
\,, \qquad
[\alpha^I_k,\tilde\alpha^J_l]  = 0 \, , \qquad
[\tilde\alpha^I_k,\tilde\alpha^J_l] = \omega_k\, \delta^{IJ}\,
\delta_{k,-l} \,.
\label{boscom}
\eeq
In addition there are the bosonic zero modes that describe the
centre of mass position $x^I_0$ and some generalised momentum $p^I_0$
with $[p_0^I,x_0^J] = - i \delta^{IJ}$. 
It is convenient to introduce the creation and annihilation operators
\beq
a^I_0 = {1\over \sqrt{2\m}} \bigl(p_0^I + i \m x_0^I\bigr) \,, \qquad
\bar{a}^I_0 = {1\over \sqrt{2\m}} \bigl(p_0^I - i \m x_0^I\bigr) \,,
\label{creani}
\eeq
in terms of which the commutation relations are then
$[\bar{a}_0^I,a_0^J] = \delta^{IJ}$. These are the familiar commutation
relations of a harmonic oscillator; this was to be expected as the
zero modes describe a point particle in a harmonic oscillator potential.
Similarly we can expand the fermionic fields in terms of modes 
$S^a_k$ and $\tilde{S}^a_k$, where
$a$ is a spinor index of $SO(8)$ and $k\in\Zop$. These modes then
satisfy the anti-commutation relations 
\beq
\{ S^a_k,S^b_l\} = \delta^{ab}\delta_{k,-l} \,,  \qquad
\{ S^a_k,\tilde{S}^b_l\}  = 0 \,, \qquad
\{\tilde{S}^a_k,\tilde{S}^b_l\}  = \delta^{ab}\delta_{k,-l} \,.
\label{fermcom}
\eeq
For future convenience we also introduce the zero-mode combinations
$\theta^a_0 = {1\over \sqrt{2}}(S^a_0 + i \tilde{S}^a_0)$ as well as  
$\bar\theta^a_0 = {1\over \sqrt{2}}(S^a_0 - i \tilde{S}^a_0)$, 
and further
\bea
\theta_R & =& \half (1 + \Pi) \theta_0 \,,  \qquad
\bar\theta_R = \half (1 + \Pi) \bar\theta_0 \,, \nn\\
\theta_L & =& \half (1 - \Pi) \theta_0 \,, \qquad
\bar\theta_L = \half (1 - \Pi) \bar\theta_0 \,.
\label{thetarl}
\eea
The theory is maximally supersymmetric: in light-cone gauge the $32$
supercharges of type IIB string theory decompose into sixteen
`kinematical' supercharges that transform in the undotted spinor
representation of SO$(8)$, and sixteen `dynamical' supercharges that
transform in the dotted representation. The kinematical supercharges
are simply given by $Q_a\equiv S^a_0$ and 
$\tilde{Q}_a\equiv \tilde{S}^a_0$, while the formula for the dynamical
supercharges is \cite{mt,bp}\footnote{We are adopting a slightly
  different normalisation for the non-zero mode contributions.} 
\begin{eqnarray}
\label{q-}
\sqrt{ 2 p^+}\, Q_{\dot a}
&=&  p_0^I \gamma^I_{\dot a b} S_0^b
   - m x_0^I \left(\gamma^I\Pi\right)_{\dot a b}
  {\tilde S}_0^b
\nonumber\\
&+& \sum_{n=1}^\infty
\left( c_n \gamma^{I}_{\dot a b}
       (\alpha_{-n}^I S_n^b + \alpha_n^I S^{b}_{-n} )
+
 \frac{{\rm i} m}{2\omega_n c_n }
 \left(\gamma^I\Pi\right)_{\dot a b}
       ({\tilde \alpha}_{-n}^I {\tilde S}_n^b
        -{\tilde \alpha}_n^I {\tilde S}^b_{-n} )
\right)\,,
\nonumber\\
\\
\label{qt-}
\sqrt{ 2 p^+}\,  {\tilde Q}_{\dot a}
&=& p_0^I \gamma^{I}_{\dot a b} {\tilde S}_0^b
  + m x_0^I\left(\gamma^I\Pi\right)_{\dot a b}{ S}_0^b
\nonumber\\
&+& \sum_{n=1}^\infty
\left(( c_n  \gamma^{I}_{\dot a b}
       ({\tilde \alpha}_{-n}^I {\tilde S}_n^b
        +{\tilde \alpha}_n^I {\tilde S}^b_{-n} )
-
 \frac{{\rm i} m}{2 \omega_n c_n }
 \left(\gamma^I\Pi\right)_{\dot a b}
       (\alpha_{-n}^I S_n^b - \alpha_n^I S^b_{-n})
\right)\,,
\nonumber\\
\end{eqnarray}
where $c_n$ is defined  by
\beq
c_n ={m \over \sqrt{2\omega_n (\omega_n -n)}}\, .
\label{cndef}
\eeq
In order to describe the anti-commutation relations of the dynamical
supercharges it is useful to introduce
$Q^\pm_{\dot a}={1\over\sqrt{2}}(Q_{\dot a} \pm i \tilde{Q}_{\dot a})$.
Then the anti-commutation relations are \cite{mt} 
$\{ Q^\pm_{\dot a}, Q^\pm_{\dot b} \} = 0$, as well as
\beq\label{dyncom}
\{ Q^+_{\dot a}, Q^-_{\dot b} \}  = 2\, \delta_{\dot a\dot b}\, H +
m\, (\gamma^{ij} \,\Pi)_{\dot a \dot b} \, J^{ij}
+ m \, (\gamma^{i'j'}\, \Pi)_{\dot a \dot b} \, J^{i'j'}  \,,
\eeq
where $J^{ij}$ are the rotation generators (see \cite{mt}) while $H$
is the light-cone Hamiltonian $H$ for the closed string in the plane-wave
background
\bea
2 \, p^+ H &=& \m \left(a^I_0\, \bar{a}^I_0
+ i\, S^a_0 \,\Pi_{ab}\, \tilde{S}^b_0 +4\right)
+ \sum_{k=1}^{\infty} \left[ \alpha^I_{-k}\alpha^I_k +
\tilde\alpha^I_{-k} \tilde\alpha^I_k
+ \omega_k \left(S^a_{-k} S^a_{k} + \tilde{S}^a_{-k} \tilde{S}^a_{k}
\right) \right]\nn\\
 &=& \m \left(a^I_0\, \bar{a}^I_0 + \theta_L^a \,\bar\theta_L^a +
\bar\theta_R^a \,\theta_R^a \right)
+  \sum_{k=1}^{\infty} \left[
\alpha^I_{-k}\alpha^I_k +
\tilde\alpha^I_{-k} \tilde\alpha^I_k
+ \omega_k \left(S^a_{-k} S^a_{k} + \tilde{S}^a_{-k} \tilde{S}^a_{k}
\right) \right]
\,.\nn\\
\label{lcham}
\eea
In the limit $m\equiv 2\pi p^+\mu \to 0$ this reduces to the usual
light-cone gauge Hamiltonian in a flat background \cite{gs1}. The
normal ordering has been chosen in (\ref{lcham}) with the
understanding that $\theta_L^a$ and $\bar\theta_R^a$ are creation
operators, while $\bar\theta_L^a$ and $\theta_R^a$ are annihilation
operators.

It is easy to see from (\ref{lcham}) that for $m\ne 0$, the ground
state of the spectrum is a single bosonic state, but the excited
states come in degenerate boson-fermion pairs.  This is consistent
with supersymmetry. In the plane-wave background, the kinematical
supercharges do not commute with the light-cone Hamiltonian, but
rather satisfy
\beq\label{hamilcom}
[H, Q_a] = - {i m \over 2 p^+} \,\Pi_{ab}\, \tilde{Q}_b \,,\qquad
[H, \tilde{Q}_a] =  {i m \over 2 p^+}\, \Pi_{ab}\, Q_b \,.
\eeq
Since the kinematical supercharges do not commute with the
Hamiltonian, we cannot use them to deduce (as one does in flat space)
that the entire space must be fermion-boson degenerate. On the other
hand, the dynamical supercharges do commute with the light-cone
Hamiltonian, but their anti-commutator (\ref{dyncom}) involves
operators on the right-hand side that annihilate the bosonic ground
state. Thus, as in the familiar situation with `unbroken
supersymmetry', the ground state is not boson-fermion degenerate, but
all excited states are.

One can also show that the full closed string spectrum of the theory
gives rise to a modular invariant torus amplitude
\cite{tak,hammou}. This is a consequence of modular identities that
are very similar to those that will be discussed in the construction
of D-branes below.

\section{D-branes in the plane wave background}

Given that we have an exact world-sheet description of the theory (at
least in light-cone gauge) it is interesting to use it to give an
exact world-sheet description of D-branes. In particular, we can
describe D-branes in terms of (i) the open strings that begin and end
on them; and (ii) in terms of the closed string boundary states that
describe the coupling of the D-branes to the closed string excitations
of the theory. Finally, we should expect that the results tie in with
the corresponding supergravity analysis.

The boundary states are coherent states in the closed string theory
that implement the effect of the corresponding D-brane. It is
convenient to describe them in the usual light cone
gauge in which $x^+=2\pi p^+\tau$. The boundary states are then
necessarily instantonic, {\it i.e.} they satisfy a Dirichlet boundary
condition in the $x^+$ direction (as well as in $x^-$).\footnote{In
order to construct boundary states for time-like branes one has to
choose a different light-cone gauge; this is briefly discussed in
\cite{bgg}. Most properties of branes are however independent of
whether the two light-cone directions are Dirichlet or Neumann.} 
In order to relate these boundary states to open strings one
considers the cylinder diagram that describes the closed string
exchange between two boundary states. This diagram also has a dual
interpretation in terms of an open string 1-loop amplitude, and these
two point of views are related to one another by exchanging the roles
of the space and time-parameters on the world-sheet. [Since the length
of the cylinder is $\Delta X^+$ while its circumference is $2\pi p^+$,
this transformation also exchanges the roles of $\Delta X^+$ and 
$2\pi p^+$. As a consequence, the mass parameter of the 
resulting open string
light cone gauge is $\hm=mt$, where $t=\Delta X^+/2\pi p^+$ is the
modular parameter of the cylinder. This is explained in detail in 
\cite{bgg,gg1}.] The open string 1-loop amplitude is simply a trace,
and thus must be an integer linear combination of characters. This is
the content of the so-called  {\it open-closed string duality
relation}.  

A significant amount of work has been done on the construction of
D-branes for the maximally supersymmetric plane-wave background. The
open string point of view has been worked out (among others) in 
\cite{dp,st,bpz,st2,gg1}, while the boundary states were constructed
in \cite{bp,bgg,gg1,st3,ggss}, and the open-closed duality relation
relating the two was studied in \cite{bgg,gg1}. The supergravity
analysis was performed, among others, in \cite{st,bmz,hy}. 

In the following we shall mainly discuss the half-supersymmetric
D-branes. Before describing the results in detail, there is one basic
point that should be stressed. As we have mentioned before, because of
the fermion mass term, the background is actually only invariant under
the SO$(4)\times$SO$(4)$ subgroup of the transverse SO$(8)$
symmetry. Unlike the situation in flat space, the D-branes are
therefore not only characterised by the dimension of their
world-volume, but also by the orientation of the world-volume relative
to the background RR flux. [This is to say, while any two flat
submanifolds of the same dimension can be mapped into one another by a
rotation in SO$(8)$, this is in general not possible by a rotation in 
SO$(4)\times$SO$(4)$.] 

For any D-brane of a given orientation, let us define
\beq
M = \prod_{I\in{\cal N}} \gamma^I \,,
\eeq
where ${\cal N}$ denotes an orthonormal basis for the Neumann
directions of the brane. Then the nature of the brane will essentially
only depend on the matrix \cite{st2,gg1}
\beq
\left(\Pi M \Pi M \right)_{ab}\,.
\eeq
This matrix encodes the relevant information about the orientation of
the brane relative to the background RR flux. There are two cases to
consider: 
\smallskip

\noindent {\bf Class I}: The first class is the one that was studied 
in \cite{bp,dp,bgg} and arises when the matrix $M_{ab}$ satisfies
\beq
\left(\Pi M \Pi M\right)_{ab} = -\delta_{ab}\,.
\label{classone}
\eeq
In this case the gluing conditions of the boundary state are the
standard bosonic conditions
\beq
\left( \alpha^I_k \mp \tilde\alpha^I_{-k} \right)
|\!|D,\eta\,\rangle\!\rangle  = 0\,, \qquad k\in\Zop\,, 
\eeq
where for a Dirichlet (Neumann) direction $I$, the upper (lower) sign
applies. The fermionic gluing conditions are simply \cite{bp}
\beq
\left( S^a_k +i\eta M_{ab} \tilde{S}^b_{-k} \right)
|\!|D,\eta\,\rangle\!\rangle  = 0\,, \qquad k\in\Zop\,, 
\eeq
where $\eta=\pm$ distinguishes a brane from an anti-brane. 
In terms of the open string boundary conditions this last identity
becomes 
\beq
S^a = \eta M_{ab} \tilde{S}^b \,, \qquad \hbox{at $\sigma=0,\pi$,}
\eeq
together with the standard boundary conditions for the bosons
\cite{dp,st}. These branes preserve half the kinematical
supersymmetries, and half the dynamical supersymmetries provided that
the D-brane is at the origin in transverse space.\footnote{It was
observed in \cite{st2,st3} that these D-branes preserve (modified)
dynamical supersymmetries even if they are not at the origin in
transverse space. The analysis of  \cite{fst} however suggests that
these modified supersymmetries are an artefact of the free theory, and
do not survive interactions.}

If the brane is oriented in such a way that it has $r$ Neumann
directions among the first four (transverse) coordinates, as well as
$s$ Neumann directions among the second four (transverse) coordinates,
it is useful to denote it as D$(r,s)$ \cite{st}. It is not difficult to show
that a D$(r,s)$ brane is of class I if $|r-s|=2$. However, there also
exist class I branes that are not of this type; the simplest example
is the oblique D5-brane of \cite{hy} (\ie\ a brane with four Neumann
directions in transverse space) that was explicitly constructed
in string theory in \cite{ggss}. 

These branes are the obvious analogous of the usual flat space branes,
and thus the open-closed duality relation should work as in flat
space. There it is a consequence of the fact that the amplitudes can
be expressed in terms of the functions \cite{polcai}
\bea
f_1(q) & =& q^{{1\over 24}} \prod_{n=1}^{\infty} (1-q^n) \,, \nn\\
f_2(q) & = &\sqrt{2}\,
 q^{{1\over 24}} \prod_{n=1}^{\infty} (1+q^n) \,,\nn\\
f_3(q) & =& q^{-{1\over 48}} \prod_{n=1}^{\infty}
\Bigl(1+q^{(n-1/2)}\Bigr) \,,\nn\\
f_4(q) & =& q^{-{1\over 48}} \prod_{n=1}^{\infty}
\Bigl(1-q^{(n-1/2)}\Bigr) \,,
\label{ffunc}
\eea
where $q=e^{-2\pi t}$. The duality transformation that relates the
open and closed point of 
view is precisely the S-modular transformation under which
$t\mapsto\tilde{t}=1/t$. Writing $\tilde{q}=e^{-2\pi\tilde{t}}$, the
above functions transform as 
\beq
f_1(q) = t^{-{1\over 2}} f_1(\tilde{q})  \,, \qquad
f_2(q) = f_4(\tilde{q}) \,, \qquad f_3(q) = f_3(\tilde{q}) \,.
\label{fmode}
\eeq
These (simple) transformation formulae guarantee that the cylinder
amplitudes transform appropriately, and that the open-closed duality
relation is satisfied. For class I branes, the relevant amplitudes
involve instead of the $f_i$ functions, 
\bea
f_1^{(\m)}(q) & =& q^{-\Delta_\m} (1-q^\m)^{{1\over 2}}
\prod_{n=1}^{\infty} \left(1 - q^{\sqrt{\m^2+n^2}}\right) \,,
\label{fdef}\\
f_2^{(\m)}(q) & = &q^{- \Delta_\m} (1+q^\m)^{{1\over 2}}
\prod_{n=1}^{\infty} \left(1 + q^{\sqrt{\m^2+n^2}}\right) \,,
\label{f2def}\\
f_3^{(\m)}(q) & =& q^{-\Delta^\prime_\m}
\prod_{n=1}^{\infty} \left(1 + q^{\sqrt{\m^2+(n-1/2)^2}}\right) \,,
\label{f3def}\\
f_4^{(\m)}(q) & =& q^{-\Delta^\prime_\m}
\prod_{n=1}^{\infty} \left(1 - q^{\sqrt{\m^2+(n-1/2)^2}}\right) \,,
\label{f4def}
\eea
where $\Delta_\m$ and $\Delta^\prime_\m$ are given as  
\bea
\Delta_\m & =& -{1\over (2\pi)^2} \sum_{p=1}^{\infty}
\int_0^\infty ds \, e^{-p^2 s} e^{-\pi^2 \m^2 / s} \,, \cr
\Delta^\prime_\m & =& -{1\over (2\pi)^2} \sum_{p=1}^{\infty}
(-1)^p \int_0^\infty ds \, e^{-p^2 s} e^{-\pi^2 \m^2 / s}\,.
\label{Deltadef}
\eea
The quantities $\Delta_m$ and $\Delta^\prime_m$ are the Casimir
energies of a single (two-dimensional) boson of mass $\m$ on a
cylindrical world-sheet with periodic and anti-periodic boundary
conditions, respectively. For $\m=0$, $\Delta_\m$ and
$\Delta^\prime_\m$ simplify to the usual flat-space values,
\bea
\Delta_0 & =& -{1\over (2\pi)^2} \sum_{p=1}^{\infty} {1\over p^{2}} =
- {1\over 24} \,, \cr
\Delta^\prime_0 & =& -{1\over (2\pi)^2}
\sum_{p=1}^{\infty} {(-1)^p \over p^{2}} =  {1\over 48}\,.
\label{mzero}
\eea
Thus $f_2^{(\m)}(q)$, $f_3^{(\m)}(q)$ and $f_4^{(\m)}(q)$ simply
reduce to the standard $f_2(q)$, $f_3(q)$ and $f_4(q)$ functions 
\cite{polcai} as $\m\rightarrow 0$.

For the case of the class I branes the open-closed consistency
condition is then a consequence of the remarkable transformation
properties of these functions \cite{bgg} (see also \cite{salitz})
\beq
f_1^{(\m)}(q) = f_1^{(\hm)}(\tq)\,, \qquad f_2^{(\m)}(q) =
f_4^{(\hm)}(\tq)\,, \qquad f_3^{(\m)}(q) = f_3^{(\hm)}(\tq)\,,
\label{beautiful}
\eeq
where $\tq=e^{-2\pi \ttt}=e^{-2\pi/t}$ and $\hm=mt$. 
In the limit $\m\rightarrow 0$ the second and third equations in
(\ref{beautiful}) reproduce the above identities for $f_2$, $f_3$ and
$f_4$. 
The identity for $f_1$ (or $\eta$) can also be derived from the first
equation of (\ref{beautiful}). In fact, both sides of
the first equation tend to zero as $\m\rightarrow 0$ since
$(1-q^\m)^{1\over 2} =\sqrt{2\pi t \m} + {\cal O}(\m)$ and
$(1-\tilde{q}^{\,\widehat{\m}})^{1\over 2}=\sqrt{2\pi \m}+{\cal O}(\m)$. 
Thus, after dividing the first equation by $\sqrt{\m}$, the limit
$\m\rightarrow 0$ becomes
\beq
f_1(q) = t^{-\half} f_1(\tilde{q}) \,,
\label{etatrans}
\eeq
thus reproducing the standard modular transformation property of the
$f_1$ (or $\eta$) function. 
\smallskip

\noindent {\bf Class II}: The second class arises when  the matrix
$M_{ab}$ satisfies
\beq
\left(\Pi M \Pi M\right)_{ab} = \delta_{ab}\, ,
\label{classtwo}
\eeq
a possibility that was not considered in \cite{bp,dp,bgg} but
arose in the supergravity analyses of \cite{st,bmz} and was later
analysed in detail in \cite{gg1} (the open string description was
independently worked out in \cite{st2}). The only class II D-branes
that preserve half the dynamical supersymmetries are the $(r,s)=(0,0)$
brane and the $(r,s)=(4,0)$ or $(r,s)=(0,4)$ brane (provided that the
Neumann boundary conditions for the bosons are modified). Neither of
these branes preserves any kinematical supersymmetries, but now the
dynamical supersymmetries are also preserved if the brane is moved
away from the origin in transverse space. For the case of the
D-instanton (or time-like D1-brane) the gluing conditions for the
bosons are as above (all transverse directions are simply Dirichlet),
but the fermionic gluing conditions are now 
\bea
\left(S^a_0 + i \eta \tilde{S}^a_0 \right) 
|\!| (0,0),\eta\,\rangle\!\rangle & = & 0  \nn \,,\\
\left(S^a_n + i \eta\, R_n^{ab}\,
 \tilde{S}^b_{-n} \right)
|\!| (0,0),\eta\,\rangle\!\rangle & = & 0  \,,\label{fermions1}
\eea
where $R_n$ is the matrix
\beq\label{rdef}
R_n = {1\over n} \left( \omega_n \bbbone - \eta\, m \Pi
\right) \,.
\eeq
The corresponding open string boundary condition
that corresponds to this gluing condition is simply 
\beq\label{dinstoub}
S=\eta \tilde{S} \,, \qquad \hbox{at $\sigma=0,\pi$.}
\eeq 
If we define the eigencomponents of $\Pi$ by 
\beq
\Pi S^\pm_n = \pm S^\pm_n \qquad 
\Pi \tilde{S}^\pm_n = \pm \tilde{S}^\pm_n \,,
\eeq
then the above gluing condition is simply 
\beq
\left( S^\pm_n + i \eta R^\pm_n \tilde{S}^\pm_{-n} \right) 
|\!| (0,0),\eta\,\rangle\!\rangle  =  0 \,,
\eeq
where
\beq
R^\pm_n = \sqrt{{\omega_n\mp \eta m \over \omega_n \pm \eta m}} \,.
\eeq
It is then straightforward to calculate the cylinder amplitude
involving two such boundary states, and one finds that it involves (if
the two boundary states have opposite $\eta$) the function
\beq
g_2^{(m)}(q) = 4\, \pi\, m\, q^{-2\Delta_m}\, q^{m/2} \,
\prod_{n=1}^{\infty} 
\left( 1 + \left({\omega_n + m \over \omega_n - m} \right) 
q^{\omega_n} \right) \, 
\left( 1 + \left({\omega_n - m \over \omega_n + m} \right) 
q^{\omega_n} \right) \,.
\eeq
Under the modular transformation $q\mapsto \tilde{q}$, this function
becomes \cite{gg1}
\beq\label{stunning}
g_2^{(m)}(q) = \hat{g}_4^{(\hm)}(\tilde{q}) \,,
\eeq
where
\beq
\hat{g}_4^{(\hm)}(\tilde{q}) = \tilde{q}^{-\widehat{\Delta}_{\hm}} \,
\prod_{l\in{\cal P}_+} (1-\tilde{q}^{\hat\omega_l})^{1/2} \,
\prod_{l\in{\cal P}_-} (1-\tilde{q}^{\hat\omega_l})^{1/2}\,,
\eeq
and $l\in{\cal P}_\pm$ provided that $l$ satisfies the
transcendental equation 
\beq
{l\pm i \hm \over l \mp i \hm} = - e^{2\pi i l} \,.
\eeq
As is explained in detail in \cite{gg1}, this then reproduces
precisely the open string 1-loop amplitude for an open string with
boundary conditions (\ref{dinstoub}) at the two ends (where the two
values of $\eta$ are different). As an aside we mention that there 
exist various generalisations of (\ref{stunning}) that were also
proven in \cite{gg1}. 
\vspace{0.5cm}

There are also supersymmetric D-branes that are neither class I nor
class II. For example the oblique D3-brane that was first discussed
from a supergravity point of view in \cite{hy} corresponds to the
choice 
\beq
M_3 = {1\over 2} (\gamma^1-\gamma^6)\, (\gamma^2+\gamma^5) \,,
\eeq
and therefore satisfies
\beq
\Pi M_3 \Pi M_3 = \gamma^1 \gamma^2 \gamma^5 \gamma^6 \,.
\eeq
This matrix has four eigenvectors with eigenvalue $+1$, and four
eigenvectors with eigenvalue $-1$. The corresponding brane is
therefore half way between being class I and class II. As is explained
in detail in \cite{ggss} it preserves a quarter of the kinematical and
dynamical supersymmetries. The open-closed duality relations are
satisfied by arguments that are a combination of the arguments that
arise for class I and class II branes.
\vspace{0.5cm}

Finally, there are also supersymmetric D-branes whose world-volume is
not flat. For example, there is a curved D7-brane whose world-volume
is described by the equation \cite{hy}
\beq\label{curved}
\sum_{i=1}^4 Z_i Z_i = c \,, \qquad Z_i = X_i + i X_{i+4} \,.
\eeq
It was shown in \cite{ggss} that the corresponding brane actually
preserves two dynamical supersymmetries. In order to describe the
relevant boundary conditions for the fermions in the open string we
introduce the space-dependent matrix
\beq
M_7 = {1\over x^I x^I}
         \left( x^i \gamma^i - x^{i+4} \gamma^{i+4} \right)
         \left( x^i \gamma^{i+4} + x^{i+4} \gamma^i \right) \,.
\eeq
The boundary condition for the fermions is then simply given by 
\beq
S = \eta M_7 \tilde{S}\,, \qquad \hbox{at $\sigma=0,\pi$.}
\eeq
The boundary condition for the bosons on the other hand receives a 
correction term due to the fact that $M_7$ depends on $Z^i$. In the
above complex basis, the relevant boundary condition is then
\cite{ggss}
\beq
\left(
P^{\bar{\jmath}i}\partial_\sigma\bar{Z}^j
-2 S \frac{\partial M_7}{\partial Z^i} \tilde{S} \right) = 0 \,,
\label{bilin} 
\eeq
where $\bar Z^j = X^j - i X^{j+4}$ and 
\beq
P^{\bar\jmath l}= \left( \delta^{jl} - {\bar Z^j Z^l \over |Z|^2}
\right) \,.
\eeq

\section{Conclusion}

In this lecture we have given a brief overview over string theory in
the maximally supersymmetric plane wave background. We have also
described some of the supersymmetric D-branes of this background. 

As has been mentioned before, the maximally supersymmetric plane-wave
background is a background with non-trivial RR flux. It would be
interesting to study such backgrounds further since they
form a `corner' of string theory that has been very little explored so
far. The simplest generalisation are other plane-wave backgrounds for 
which the coefficient of the $(dx^+)^2$ term in the metric is of the
form $A_{IJ}(x^+) x^I x^J$. The resulting world-sheet theories always  
preserve 16 kinematical supersymmetries since the fermions of
these theories are always free. On the other hand, these backgrounds
do not preserve in general any dynamical supersymmetries; if they do,
this is a consequence of some of the bosonic fields being free as
well. These backgrounds therefore only preserve dynamical
supersymmetries if they behave in some respect as the maximally
supersymmetric case. 

A more interesting class of backgrounds are therefore 
pp-wave backgrounds for which the coefficient of $(dx^+)^2$ in the
metric is not necessarily quadratic in the transverse coordinates. It
was shown in \cite{mm} (see also \cite{tr}) that there are backgrounds
of this type which preserve some dynamical supersymmetries without
involving free boson and fermion fields. Furthermore it was shown
that at least some of these do describe exact string solutions
\cite{BM}. In the most interesting examples, the relevant world-sheet  
theories are integrable. It would be very interesting to make use of
this integrable structure in order to gain insight into the nature of
these theories. Some progress has recently been made in this
direction in \cite{TF}.
\\ \\
{\bf Acknowledgement} The author would like to thank the organisers of the
symposium for organising a very pleasant conference, and for giving him
the opportunity to speak. He also thanks Michael Green for a very
enjoyable collaboration on which much of what is presented here is
based, and Tako Mattik for a careful reading of the manuscript.


\end{document}